\documentstyle[14pt,epsf,rotate]{article}
\newcommand{\Arg}{\mathop{\rm Arg}\nolimits}
\newcommand{\arctg}{\mathop{\rm arctg}\nolimits}
\newcommand{\ch}{\mathop{\rm ch}\nolimits}
\newcommand{\lsim}{\mathop{\raisebox{0.3em}{$<$}%
\hspace{-0.8em}\raisebox{-0.3em}{$\sim$}}}
\newcommand{\gsim}{\mathop{\raisebox{0.3em}{$>$}%
\hspace{-0.8em}\raisebox{-0.3em}{$\sim$}}}

\begin{document}
\author{A.~A. Slavnov \thanks{e-mail:$~~$ slavnov@mi.ras.ru} \ \ and 
\ \ N.~V. Zverev \thanks{e-mail:$~~$ zverev@mi.ras.ru} \\
{\it Steklov Mathematical Institute, Russian} \\
{\it Academy of Sciences, Gubkin st. 8, 117966, Moscow, Russia} \\
{\it Moscow State University, Physical Faculty,} \\
{\it 117234, Moscow, Russia}}
\title{Nonlocal lattice fermion models on the 2d torus}
\date{}
\maketitle

\begin{abstract}
Abelian fermion models described by the SLAC action are considered on a 
finite 2d lattice. It is shown that modification of these models by 
introducing additional Pauli -- Villars regularization supresses nonlocal 
effects and provides agreement with the continuum results in vectorial 
U(1) models. In the case of chiral fermions the phase of the determinant 
differs from the continuum one.
\end{abstract}

\newpage

\section {Introduction}

In a recent paper \cite{SlZv} we considered lattice fermion models on the 
2d lattice with Wilson action improved by Pauli -- Villars (PV) 
regularization. It was shown that in spite of the chiral symmetry breaking 
for finite lattice spacing $a$, its effects are supressed by PV 
regularization and the model provides a good agreement with the continuum 
results both in the perturbative and nonperturbative region. Nevertheless 
lack of chiral symmetry for finite $a$ is not quite harmless as to get a
good agreement for large external fields one needs big lattices.

So it would be highly desirable to have a formulation which preserves 
chiral symmetry for a finite lattice spacing as well.

"No-go" theorem \cite{NiNe} forbids any local formulation preserving chiral 
symmetry, therefore we consider a nonlocal SLAC model proposed originally 
in paper \cite{DWY}. It is known that the original formulation of the model 
requires introduction of nonlocal counterterms \cite{KS1, KS2}, which 
makes it unacceptable for practical calculations. However the model can be 
improved in the same way as it has been done for Wilson action by 
introducing additional gauge invariant PV regularization which supresses 
the contribution of momenta close to the edge of the Brillouin zone. It 
was shown \cite{Sl} that for anomaly free models on an infinite lattice in 
the framework of perturbation theory all nonlocal effects can be supressed 
in this way and one gets a manifestly chiral invariant formulation of 
anomaly free models.

The propose of this paper is to check these results both perturbatively 
and nonperturbatively for the models on the 2d finite lattice. We found 
that in the case of vectorial models this approach works leading to a 
reasonable agreement with continuum results. However for chiral fermions, 
even when perturbative anomaly is absent, we observed a discrepancy in the 
value of the phase of continuum and lattice determinants. The origin of 
this discrepancy is discussed.

\newpage

\section {Vectorial lattice model}

In this section we consider the vectorial model described by the action
\cite{DWY}:
\begin{equation}\label{1}
I_{VS} = \sum_{x, y, \mu} \overline{\psi}(x)\gamma_\mu {\cal D}_\mu(x-y)
\exp\Bigl[{\rm i} \sum_{z_\mu=x_\mu}^{y_\mu} A_\mu(z)\Bigr] \psi(y). 
\end{equation} 
Here $-N/2+1 \le x_\mu \le N/2$, $\mu=0, 1$. The lattice spacing is chosen 
to be equal to 1. We restrict ourselves by the case of a constant external 
fields
$$
A_\mu=\frac{2\pi}{N}h_\mu, \qquad h_\mu={\rm const}.
$$
The Fermi field $\psi$ satisfies antiperiodic boundary conditions. 
${\cal D}_\mu(x)$ is the lattice (SLAC) derivative:
$$
{\cal D}_\mu(x)=\frac{1}{N^2}\sum_{p=-N/2+1}^{N/2}{\rm i}{\cal P}_\mu(p)
\exp\frac{2\pi{\rm i}}{N}\biggl(p-\frac{1}{2}\biggr)x,
$$
where
$$
{\cal P}_\mu(p)=\frac{2\pi}{N}\biggl(p_\mu-\frac{1}{2}\biggr), \quad
-N/2+1 \le p_\mu-mN \le N/2, \quad m=0, \pm 1, \dots.
$$
The action (\ref{1}) is gauge invariant but it is not local.

A straightforward calculation gives for the vectorial fermion determinant
normalized to 1 at $h=0$, the following expression:
\begin{equation}\label{2}
D_{VS} = \prod_{p=-N/2+1}^{N/2} \frac{B^2(p, h)}{B^2(p, 0)},
\end{equation}
where $B^2(p, h) = \sum\limits_{\mu=0}^{1} B^2_\mu(p, h)$, $B_\mu$ is the 
Fourier transform of the covariant derivative
\begin{equation}\label{3}
B_\mu(p, h) = \frac{\pi}{N} \sum_{\scriptstyle z=-N/2+1 \atop \scriptstyle
z \neq 0}^{N/2}(-1)^{z+1}\frac{\sin\frac{2\pi}{N}\left( p_\mu-h_\mu-
\frac{1}{2}\right)z}{\sin\frac{\pi z}{N}}.
\end{equation}
The dependence of $B_\mu / 2\pi$ on $(p_\mu-h_\mu-1/2)/N$ is presented at 
Fig.1. One sees that on a finite lattice, the saw-tooth form of the SLAC 
derivative is smoothen and contrary to the infinite lattice case may be 
approximated by a continuous curve. This fact will be important for 
further discussion.

\begin{figure}[th]
\epsfxsize=16cm

\vspace*{-7mm}
\hspace*{-3mm}
\rotate[r]{\epsfbox{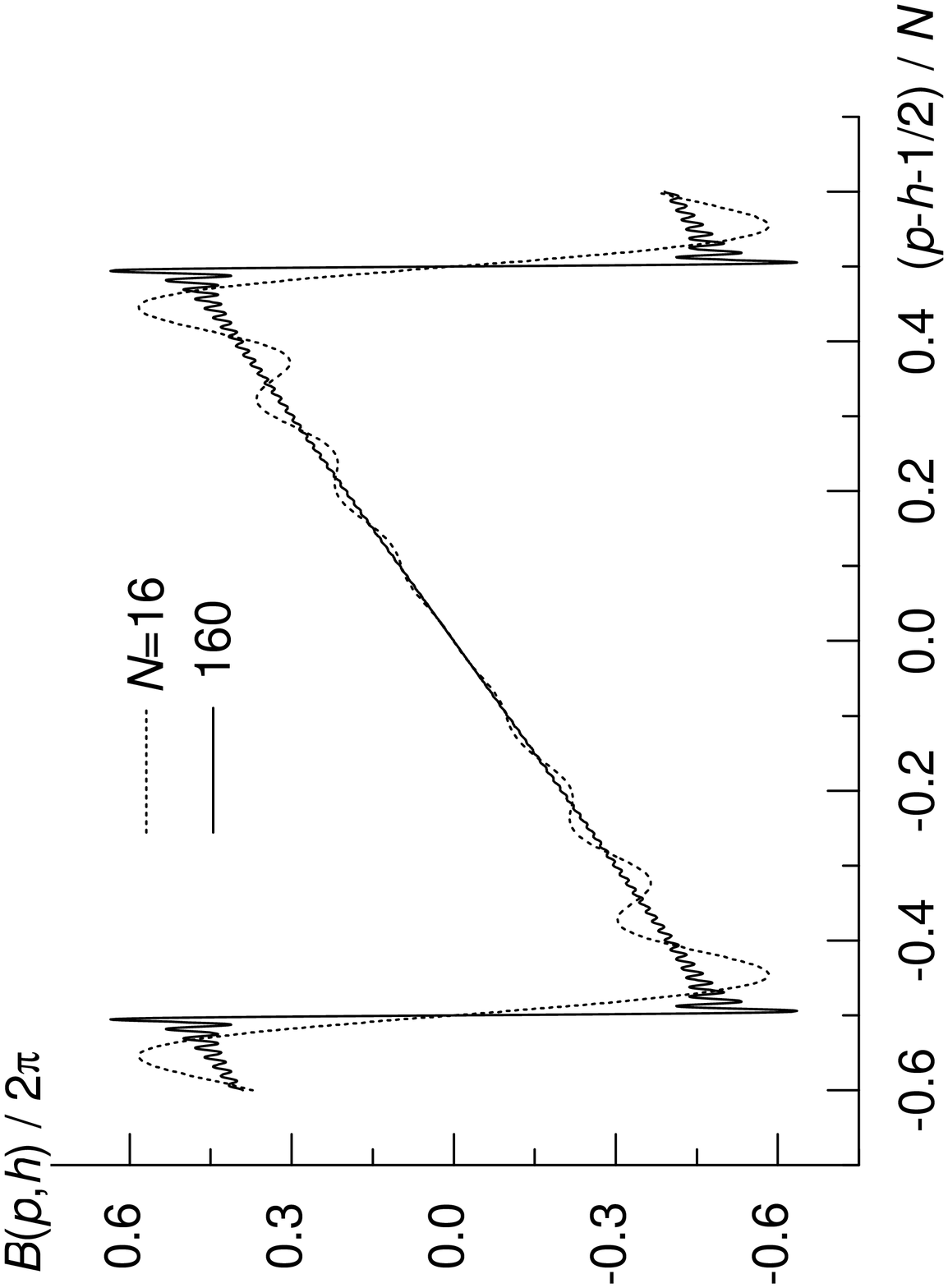}}

\vspace*{-48mm}
{\bf Fig.1.}
Covariant derivative $B(p, h)/2\pi$ as function of $(p-h-1/2)/N$ \\ 
\hspace*{16mm} at $N\!=\!16$ and 160
\bigskip
\end{figure}

The lattice determinant (\ref{2}) has to be compared with the known 
result for the continuum theory \cite{AGMV, NaNe, FoRD}
\begin{equation}\label{4}
D_{VC} = {\rm e}^{-2\pi h_1^2}\prod_{n=1}^{\infty} \Bigl| F[n, h] F[n, -h] 
\Bigr|^2,
\end{equation}
where
$$
F[n, h]=\frac{{\displaystyle 1 + {\rm e}^{-2\pi(n-1/2)+2\pi{\rm i}
(h_0+{\rm i}h_1)}}} {{\displaystyle 1 + {{\rm e}^{-2\pi(n-1/2)}}}}.
$$

The determinants (\ref{2}), (\ref{4}) satisfy the following symmetry 
properties:
$$
D[h_0, h_1] = D[h_1, h_0] = D[-h_0, h_1] = D[h_0+n_0, h_1+n_1], 
$$
where $n_0, n_1 =0, \pm 1, \pm 2, \dots$. Therefore it is sufficient to 
consider the fields $h_\mu$ only in the range $0 \le h_0 \le h_1 \le 1/2$.

\begin{figure}[th]
\epsfxsize=16cm

\vspace*{-7mm}
\hspace*{-3mm}
\rotate[r]{\epsfbox{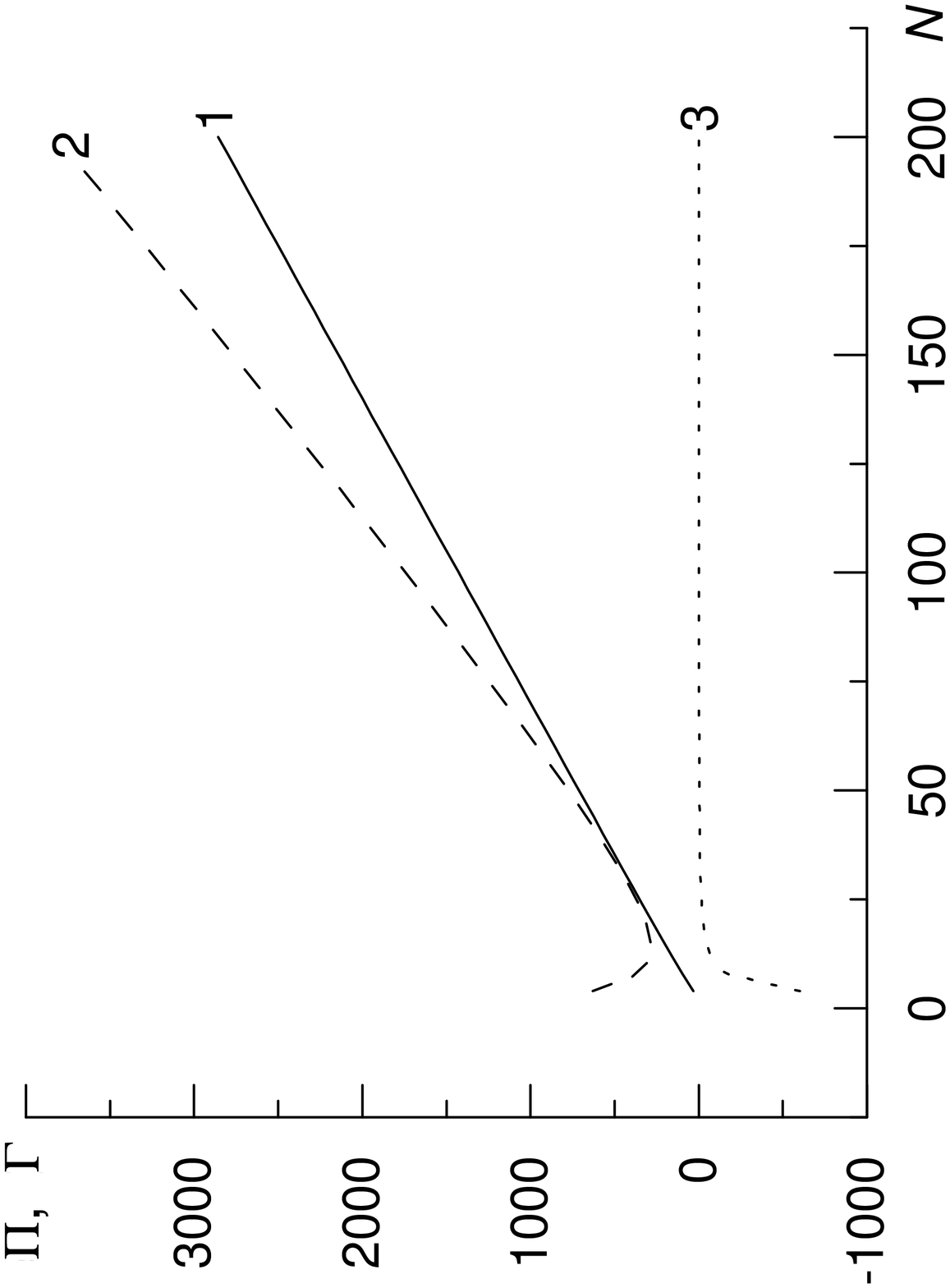}}

\vspace*{-48mm}
{\bf Fig.2.}
Vectorial diagrams as functions of $N$ at external momentum $k\!=\!0$: \\
\hspace*{16mm} 1 -- $\Pi_{VS}(0)$;   \qquad  2, 3 -- with four lines: \quad
2 -- $\Gamma_{0000}(0)$,  \quad 3 -- $\Gamma_{0101}(0)$
\bigskip
\end{figure}

Using the eq. (\ref{2}) one can easily write the expression for the diagrams
with two and more external $h$-lines. It looks as follows:
\begin{eqnarray}\label{5}
\frac{\partial^2}{\partial h_\mu \partial h_\nu} \ln D_{VS} = \sum_
{p=-N/2+1}^{N/2} \biggl\{ 2\delta_{\mu\nu} \cdot \frac{B'^2_\mu(p, 
h)+ B_\mu(p, h) B''_\mu(p, h)}{B^2(p, h)} - \biggr. \nonumber \\
\biggl. - \frac{4 B_\mu(p, h)B'_\mu(p, h) B_\nu(p, h)B'_\nu(p, h)}
{\left[B^2(p, h)\right]^2}\biggr\},
\end{eqnarray} 
where $'$ means a derivative with respect to $h$.
We wish to study the behaviour of different diagrams at $N\!\to\!\infty$.
Computer simulations of the second and fourth order diagrams as functions of
$N$ were performed using eqs (\ref{5}) and (\ref{3}). The results are 
presented at Fig.2. One sees that the polarization operator $\Pi_{VS}(0)$ 
and the diagram with four external lines $\Gamma_{0000}(0)$ diverge like $N$. 
These calculations show that the SLAC action does not provide the correct 
continuum limit. A mass renormalization which one could expect on the basis
of power counting in the continuum theory does not save the situation as the 
diagrams with more than two external lines diverge in the limit 
$N\!\to\!\infty$.

We shall try to improve the model by introducing additional PV 
regularization according to \cite{Sl}. The regularized action looks as 
follows:
$$
I_{VR} = I_{VS} + I_{PV},
$$
\begin{eqnarray*}
I_{PV} = \sum_{r} \biggl\{ \sum_{x, y, \mu} \overline{\phi}_r(x)\gamma_\mu 
{\cal D}_\mu(x-y) \exp\Bigl[{\rm i} \sum_{z_\mu=x_\mu}^{y_\mu} A_\mu(z)
\Bigr] \phi_r(y) + \biggr. \\
\biggl. + \sum_{x} M_r \overline{\phi}_r(x)\phi_r(x) \biggr\}.
\end{eqnarray*}
Here $\phi_r$ are Bose and Fermi PV fields having the same spinorial and 
internal structure as $\psi$.

The estimates of asymptotic behaviour of different diagrams, analogous to 
the ones presented above, show that if one uses one PV field the diagrams 
have asymptotics $\sim\!M^2 N^2$. Therefore to supress the contribution of 
momenta close to the edge of the Brillouin zone which are responsible for 
nonlocal effects, one has to choose $M\!\ll\!\frac{1}{N}$. Such small values 
of PV field masses are not acceptable as they are comparable with masses of
physical particles. Moreover, as we shall see from numerical analysis, the 
model with one PV field is very sensitive to the particular choice of $M$ 
and therefore the results are not stable.

To get reliable results one needs to introduce at least three PV fields. In
this case the diagrams have asymptotic behaviour $M^4N^2$, and choosing
$\frac{1}{N}\! \ll\! M \!\ll\! \frac{1}{\sqrt{N}}$ one can supress the 
contribution of momenta close to the edge of the Brillouin zone.

Below we present the results of numerical calculations for the cases of 1 
and 3 PV fields.

The regularized determinant has a form
\begin{equation}\label{6}
D_{VR} = D_{VS}[h] D_{PV}[h],
\end{equation}
where $D_{VS}[h]$ is given by eq. (\ref{2}) and $D_{PV}$ is defined as 
follows:
$$
D_{PV} = \prod_{r}\prod_{p=-N/2+1}^{N/2} \biggl( \frac{B^2(p, h)+M_r^2}
{B^2(p, 0)+M_r^2} \biggr)^{c_r}.
$$
Here $c_r\!=\!1$ or $-1$ corresponds to the case of Fermi or Bose PV field.

\begin{figure}[th]
\epsfxsize=16cm

\vspace*{-7mm}
\hspace*{-3mm}
\rotate[r]{\epsfbox{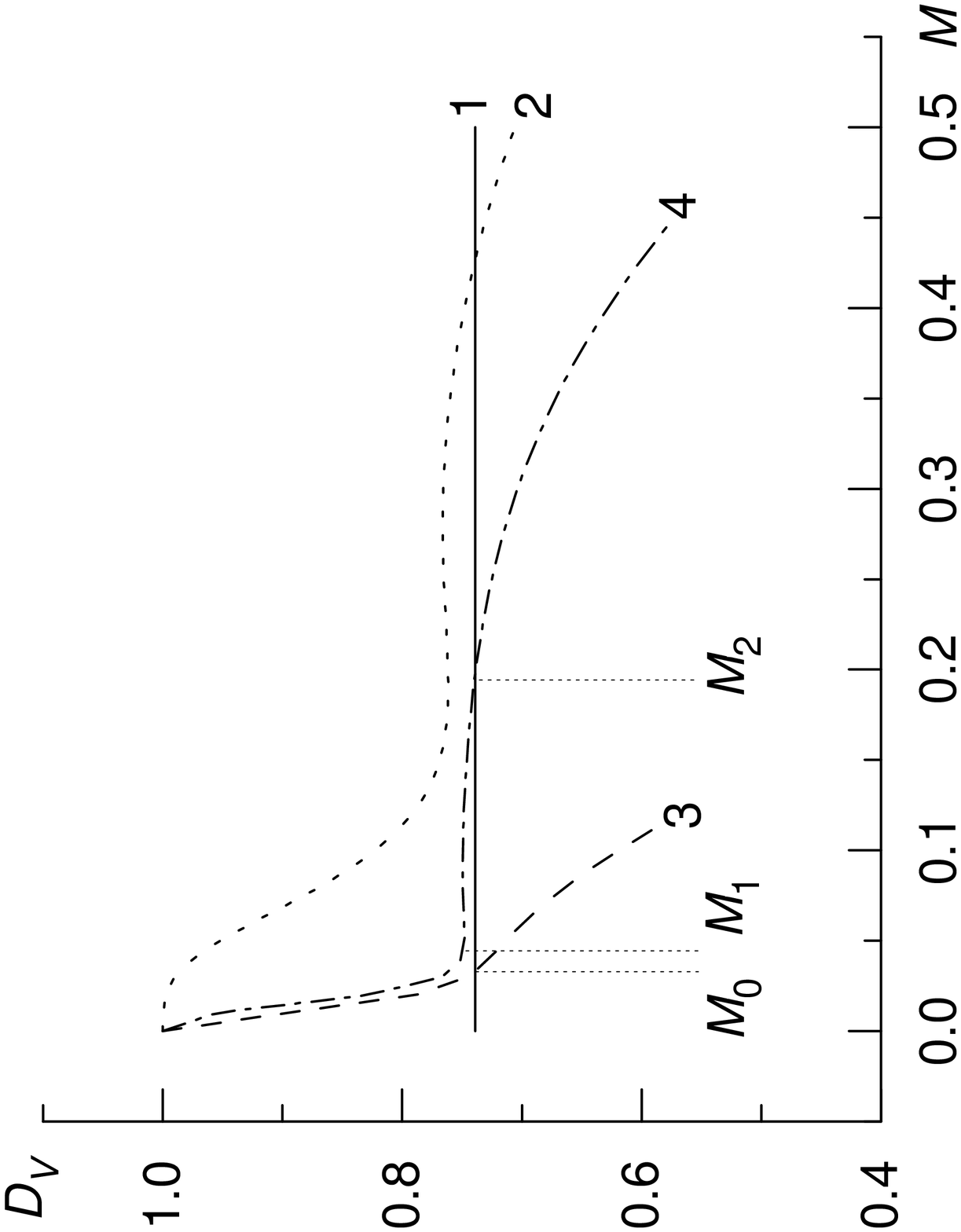}}

\vspace*{-48mm}
{\bf Fig.3.}
Vectorial determinants $D_V$ as functions of $M$ at $h_0\!=\!h_1\!=\!0.2$: \\
\hspace*{16mm} 1 -- $D_{VC}$ on the torus; \qquad  
2, 3, 4 -- $D_{VR}$ with PV fields: \\
\hspace*{16mm} 2 -- $N\!=\!32$, 3 PV fields; \qquad  3, 4 -- $N\!=\!160$: \\
\hspace*{16mm} 3 -- 1 PV field, \qquad 4 -- 3 PV fields
\bigskip
\end{figure}

The results of calculations are presented at Fig.3. One sees that in the 
case of one PV fields the agreement is achieved for $N$=160 only at one 
particular value of $M_0$=0.026 -- 0.033. Even small variation in the 
value of $M$ leads to a large discrepancy between lattice and continuum 
results. In the case of 3 PV fields the lattice results agree with the 
continuum in the rather big interval of values of PV fields masses. For 
$N$=160 the agreement is observed for $M_1 \!=\! 0.04 \!\le\! M \!\le\! M_2 
\!=\! 0.2$. These values are practically independent of $h$, except for the 
case $h_\mu \! \to \! 0.5$. Comparing the data at $N$=32 and $N$=160 one 
notes that if $N$ grows the mass values decrease and the minimal value of 
$M_1$ at 3 PV fields behaves like $N^{-3/4}$.

Let us discuss the behaviour of the lattice determinant in the case of 3 PV
fields in some more detailes. Due to the fact that the covariant derivative 
$B_\mu(p, h)$ sharply goes to zero at the border of the Brillouin zone, it 
is instructive to separate the products in eqs (\ref{2}), (\ref{7}) into 
three parts:
\begin{equation}\label{7}
D_{VR}=D_{\rm in} D_b D_a.
\end{equation}
Here $D_{\rm in}$ is a product over $p$ excluding external points, $D_b$ is 
the product over $p$ belonging to the edges of the Brillouin zone and $D_a$ 
is the product of $p$ corresponding to the vertices of the zone.

Let us start with $D_{\rm in}$. According to the behaviour of $B_\mu(p, h)$
(see Fig.1), in this region difference between regularized diagrams and 
the continuous ones is of order $1/MN$. Near the border of this region 
$B_\mu(p, h)$ and its derivatives are of order 1. It allows to expand the 
corresponding terms in regularized diagrams over $M^2$. In the case of 3 PV 
fields the first nonvanishing term is $\sim\! M^4$ leading to the asimptotic 
behaviour $\sim \! M^4 N^2$. Summarizing these estimates we get
\begin{equation}\label{8}
D_{VR} = D_{VC}\left(1 + {\rm O}\left(1/MN\right) + {\rm O}(M^4 N^2)
\right), \qquad N \to \infty, 
\end{equation}
when $\frac{1}{N} \ll M \ll \frac{1}{\sqrt{N}}$.

For the further estimates of values $D_a$ and $D_b$, using numerical data
presented at Fig.1, we approximate the covariant derivative $B_\mu(p, h)$ 
by the following two lines (ignoring oscillations):
$$
B_\mu(p, h)\! \approx \! \frac{2\pi}{N}\bigl(p_\mu - h_\mu - 1/2 \bigr), 
\qquad p_\mu = -N/2+2,\dots,N/2-1;
$$
\begin{equation}\label{9}
B_\mu(p+N/2, h)\! \approx\! -2\pi\bigl(p_\mu - h_\mu - 1/2 \bigr), \qquad 
p_\mu = 0,1.
\end{equation}

The value of $D_a$ at $|h_\mu|\!\le\!1/2$ has the form following from the 
eqs (\ref{6}) and (\ref{9}):
$$
D_a = D_{a0}[h_0, h_1] D_{a0}[h_0, -h_1] D_{a0}[h_1, h_0] D_{a0}[h_1, -h_0]
/ D_{a0}^4[0, 0].
$$
Here
$$
D_{a0}[h_0, h_1] = \frac{\Bigl[ \bigl(\frac{1}{2} - p\bigr)^2 + 2(M/2\pi)^2 
\Bigr] \bigl(\frac{1}{2} - p\bigr)^2}{\Bigl[ \bigl(\frac{1}{2} - p\bigr)^2 
+ (M/2\pi)^2 \Bigr]^2}, 
$$ 
where $\left(\frac{1}{2} - p\right)^2 = \sum\limits_\mu \left(\frac{1}{2} - 
p_\mu\right)^2$. One sees that at all values of $h$ in the interval $|h_\mu|
\!\le\!1/2$ except the trivial case $|h_0|\!=\!|h_1|\!=\!1/2$ when $D_{VR}\!
=\!D_{VC}\!=\!0$
\begin{equation}\label{10}
D_a \!\to\! 1 \qquad \mbox{at } M(N)\! \to\! 0.
\end{equation}

Now we consider the value $D_b$. Its expression at $|h_\mu|\!\le\!1/2$ looks
as follows:
$$
D_b = D_{b0}[h_0, h_1] D_{b0}[h_0, -h_1] D_{b0}[h_1, h_0] D_{b0}[h_1, -h_0]
/ D_{b0}^4[0, 0].
$$
Here
$$
D_{b0}[h_0, h_1] = \prod_{p_0=-N/2+2}^{N/2-1}\frac{\bigl[ G[p_0, h] + 
2(M/2\pi)^2 \bigr] G[p_0, h]}{\bigl[ G[p_0, h] + (M/2\pi)^2 \bigr]^2},
$$
where $G[p_0, h] = \frac{1}{N^2}\left(p_0 - h_0 -\frac{1}{2}\right)^2 + 
\left(\frac{1}{2} - h_1 \right)^2$.

Let us extend in the last equation from the interval $-N/2+2\!\le\! p_0\!\le\!
N/2-1$ to $-\infty\!<\! p_0\!<\!\infty$. In additional domain $\left|p_0 - 
\frac{1}{2}\right|\!\ge\!\frac{N-1}{2}$ the value $D_{b0}$ behaves like 
$$ 
\sum_{\left|p_0-\frac{1}{2}\right|\!\ge\!\frac{N-1}{2}} \ln\left(1+ 
\frac {2(M/2\pi)^2}{G[p_0, h]}\right)\lsim \left(\frac{MN}{\pi}\right)^2 
\sum_{p_0=N}^{\infty}\frac{1}{p_0^2} \lsim M^2 N.  
$$ 
Using this estimation one can transform $D_{b0}$ to the form 
$$ 
D_{b0}[h_0, h_1] = \frac{H[h, M\sqrt{2}] H[h, 0]}{H^2[h, M]}\left(1+{\rm O}
(M^2 N)\right), 
$$
where $H[h, M] = \ch 2\pi N \sqrt{\left(\frac{1}{2} - h_1 \right)^2 + \left
(\frac{M}{2\pi}\right)^2} - \cos 2\pi\left(h_0+\frac{1}{2}\right)$. \\
According to this formula, we get for the value $D_b$ at $\left|\frac{1}{2}
- h_\mu\right|\!\gg\!M$ the following expression:
\begin{eqnarray}\label{11}
D_b = \exp \biggl\{ -\frac{M^4 N}{32\pi^3} \biggl[ \bigl(1/2 - h_1 \bigr)^{-3}
 + \bigl(1/2 + h_1 \bigr)^{-3} + \biggr. \biggr. \nonumber \\ 
\biggl. \biggl. + \bigl(1/2 - h_0 \bigr)^{-3} + \bigl(1/2 + h_0 \bigr)^{-3} 
- 32 \biggr]\biggr\}.  
\end{eqnarray}

It follows from eqs (\ref{7}), (\ref{8}), (\ref{10}), (\ref{11}) that for 
$\frac{1}{N}\!\ll\! M \!\ll\!\frac{1}{\sqrt{N}}$ our regularized lattice model 
agrees with the continuum toron model for the fields $h_\mu$ in the interval 
$\Bigl|\frac{1}{2} \pm h_\mu \Bigr| \gsim (M^4N)^{1/3}$.

In the region $(M^4N)^{1/3} \gg \Bigl|\frac{1}{2} \pm h_\mu \Bigr| \gg M$, 
the lattice determinant decreases sharply when $|h_\mu|\!\to\!1/2$ and the 
agreement with the continuum model is lost. This effect is due to vanishing 
of the covariant derivative $B_\mu(p, h)$ at the border of the Brillouin 
zone, and as our calculations show it comes from the contribution of the 
edges of the Brillouin zone.

If we denote the value of $h$ at which $D_{VR}$ and $D_{VC}$ start to differ
by more than 5\% by $h^\star$, i.e when $D_{VR}/D_{VC}\!=\!\xi\!=\!0.95$, it 
follows from eq. (\ref{12}) that
$$
|h^\star_\mu|=\frac{1}{2}-\biggl( \frac{M^4N}{32 \pi^3 
\ln\xi^{-1}}\biggr)^{1/3}.  
$$
One sees that in the limit $N \!\to\! \infty$ the regularized model agrees 
with the continuum toron model in the whole interval of $h$ values.

\begin{figure}[t]
\epsfxsize=16cm

\vspace*{-7mm}
\hspace*{-3mm}
\rotate[r]{\epsfbox{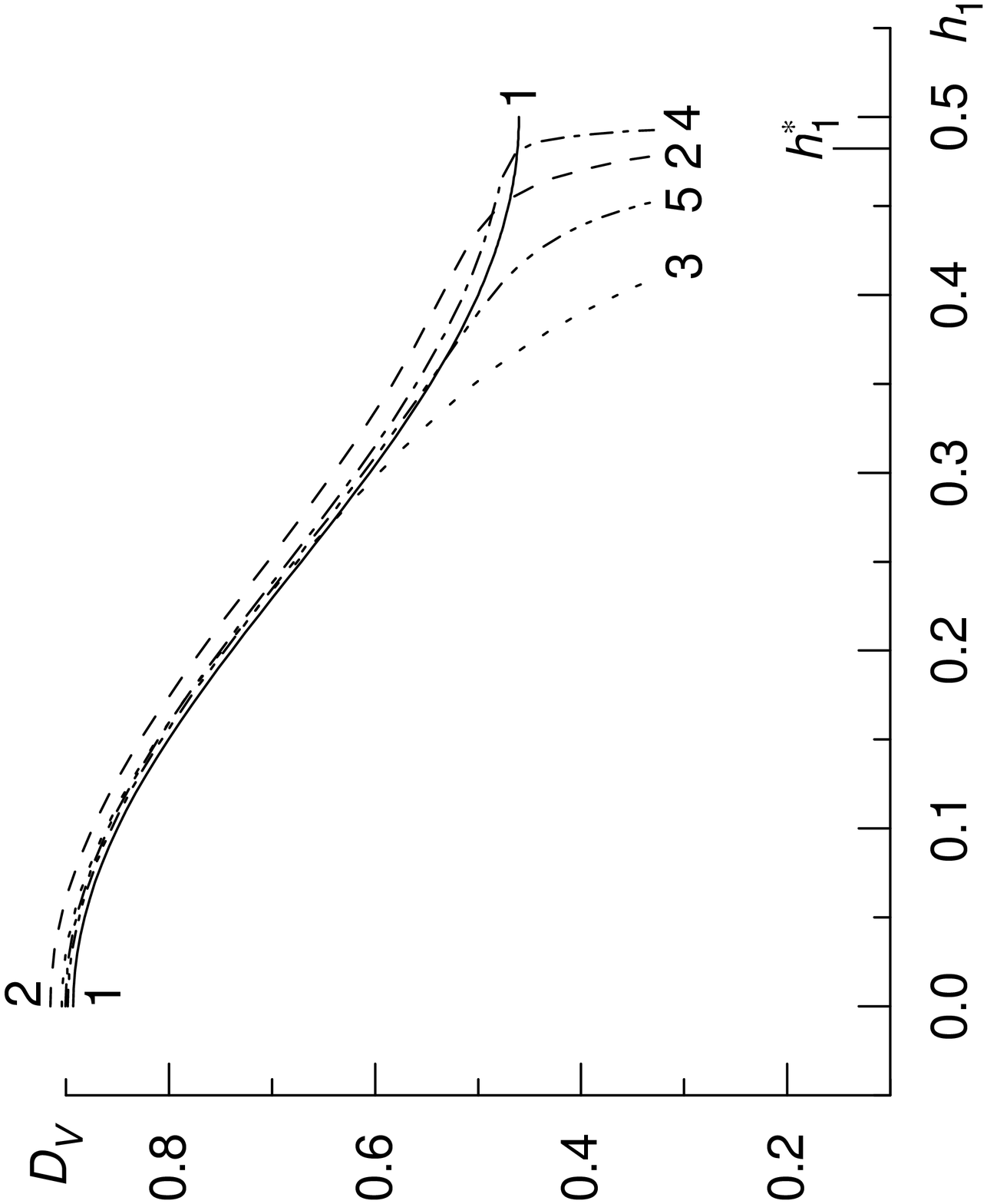}}

\vspace*{-48mm}
{\bf Fig.4.}
Vectorial determinants $D_V$ as functions of $h_1$ at $h_0=0.2$: \\
\hspace*{16mm} 1 -- $D_{VC}$ on the torus; \qquad  
2, 3, 4, 5 -- $D_{VR}$ with 3 PV fields: \\
\hspace*{16mm} 2, 3 -- $N\!=\!32$: \quad 2 -- $M\!=\!M_1\!=\!0.15$, \quad 
3 -- $M\!=\!M_2\!=\!0.4$; \\
\hspace*{16mm} 4, 5 -- $N\!=\!160$: \quad 4 -- $M\!=\!M_1\!=\!0.04$, \quad  
5 -- $M\!=\!M_2\!=\!0.15$
\bigskip
\end{figure}

These analytic results are in a good agreement with numerical calculations 
of $D_{VR}$ and $D_{VC}$ according to eqs (\ref{6}), (\ref{4}). At Fig.4 
the values of $D_{VR}$ and $D_{VC}$ calculated for different $h$ and $M$ 
are presented. One sees that for $N$=160 3 PV fields provide a good 
agreement with the continuum for $|h| \! < \! 0.4$.

\section {Phase of the lattice chiral determinant}

In this section we consider the possibility to use the SLAC action for 
regularizing chiral U(1) model. Obviously such a regularization could be
successful only for anomaly free models, as SLAC action is chiral invariant 
and therefore cannot reproduce anomaly. However our discussion in the 
previous section shows that even in the case when perturbative anomalies are 
compensated as in 11112 model, this regularization most probably fails. 
Indeed the possibility to supress the contribution of momenta close to the 
border of the Brillouin zone in anomaly free models by introducing gauge 
invariant vectorial interaction of PV fields is related to the fact that 
usually we deal with a finite number of divergent diagrams, whose sum is 
anomaly free models is parity conserving. For example in the 2d U(1) model 
with 4 positive chirality fermions with charge 1 and one negative chirality
fermion with charge 2 the sum of anomalous second order diagrams is purely 
vector like. For nonzero external momenta all higher order diagrams are 
convergent and the contribution to them of momenta close to $\pi/a$ is 
negligible. This is the reason why PV regularization was successfully applied
to the SLAC action on the infinite lattice, see \cite{Sl}. However in the 
toron model with SLAC action, as was shown in the previous section, there are 
"divergent" diagrams with more than two external lines. That means the 
contribution of boundary momenta $p_\mu \! \sim \! \frac{\pi}{a}$ is important 
not only for two-point diagram, but for other diagrams as well. The sum of the 
diagrams with more than 2 external lines is not vector-like and therefore 
cannot be supressed by vectorial PV interaction. Using PV regularization one
can easily obtain the agreement with the continuum case for the modulus of 
determinant. However vectorial PV interaction does not influence the phase of
the determinant. Therefore the SLAC action can provide a correct phase 
only if for some reasons the contribution of boundary momenta cancel by 
ifself. Moreover it is easy to show that nonzero contribution to phase comes 
from the diagrams with more than two external lines and hence anomaly 
cancelation does not help.

So it is sufficient to check the phase of a model with one positive chirality
fermion described by the standard SLAC action (\ref{1}), where one has to 
consider $\psi$ as the Weyl fermion
$$
\psi = \frac{1+\gamma_3}{2}\psi.
$$
No PV fields are needed.

A straightforward calculation gives the result
\begin{equation}\label{12}
D_{+S} = \prod_{p=-N/2+1}^{N/2}\frac{B_0(p, h)+{\rm i}B_1(p, h)}{B_0(p, 0)
+{\rm i}B_1(p, 0)},
\end{equation}
where $B_\mu(p, h)$ is the covariant derivative defined by the eq. (\ref{3})
(see also Fig.1).

The corresponding expression for the determinant in continuum theory on the 
torus looks as follows \cite{AGMV, NaNe, FoRD}:
\begin{equation}\label{13}
D_{+C} = {\rm e}^{{\rm i}\pi h_1 (h_0+{\rm i}h_1)} \prod_{n=1}^{\infty} F[n,
h] F[n, -h],
\end{equation}
where $F[n, h]$ is defined by eq (\ref{4}).

The determinants (\ref{12}), (\ref{13}) satisfy the following symmetry 
properties:
$$
D_+[h_0, h_1] = D_+^*[h_1, h_0] = D_+^*[-h_0, h_1].
$$
Due to chiral invariance of the SLAC action the lattice determinant 
(\ref{13}) is periodic in $h$:
$$
D_{+S}[h_0, h_1] = D_{+S}[h_0+n_0, h_1+n_1], \qquad n_0, n_1 =0,\pm1, 
\pm2, \dots.
$$ 
The continuum theory is anomalous and the corresponding determinant 
satisfies the condition 
$$ 
D_{+C}[h_0+n_0, h_1+n_1] = {\rm e}^{{\rm i}\pi(n_0 h_1 - n_1 h_0)} D_{+C}
[h_0, h_1].
$$ 
It follows that one can hope at most on the agreement only for the fields 
$h_\mu$ in the interval $|h_\mu|\!\le\!0.5$. It is sufficient to consider 
$0\!\le\!h_0\!\le\!h_1\!\le\!0.5$.

Let us firstly make analytic estimates of $\Arg D_{+S}$. In the previous 
section we showed that due to sharp decreasing of $B_\mu(p, h)$ near the 
border of the Brillouin zone this region gives a considerable contribution 
which may spoil the agreement with the continuum case. For that reason we 
shall study the behaviour of $\Arg D_{+S}$ in the interier and boundary part
of the zone separately. Using the notations introduced in the previous 
section we present $\Arg D_{+S}$ as a sum
\begin{equation}\label{14}
\Arg D_{+S} = \Arg D_{+\rm in} + \Arg D_{+b} + \Arg D_{+a} \bmod 2\pi.
\end{equation}
To estimate the separate terms we adopt a linear approximation (\ref{9}) of 
the covariant derivative $B_\mu(p, h)$. Eq. (\ref{12}) leads to the following
expression for different terms:
\begin{equation}\label{15}
\Arg D_{+\rm in} = \sum_{p=-N/2+2}^{N/2-1} \arctg \frac{h_0 \left(p_1 - 
\frac{1}{2} \right) - h_1 \left(p_0 - \frac{1}{2} \right)}{\sum\limits_{\mu}
\left(p_\mu - h_\mu - \frac{1}{2} \right)\left(p_\mu - \frac{1}{2} \right)}
\quad \bmod 2\pi,
\end{equation}
$$
\Arg D_{+a} = \arctg \frac{16 h_0 h_1 \left( h_0^2 - h_1^2 \right)}
{4\left(h_0^2 - h_1^2 \right)^2 - 16h_0^2 h_1^2 + 1},
$$
\begin{eqnarray*}
\Arg D_{+b} = \Arg D_{+b0}[h_0, h_1] + \Arg D_{+b0}[-h_0, -h_1] + \\
+ \Arg D_{+b0}[-h_1, h_0] + \Arg D_{+b0}[h_1, -h_0] \bmod 2\pi, 
\end{eqnarray*}
$$ \Arg D_{+b0}[h_0, h_1] = \sum_{p_0=1}^{N/2-1}\arctg \frac{2 h_0 N \left( 
h_1 - \frac{1}{2} \right)}{\left(p_0-\frac{1}{2} \right)^2 + N^2\left(h_1 - 
\frac{1}{2} \right)^2 - h_0^2} \quad \bmod 2\pi.
$$ 

Due to the symmetry properties of the determinant $D_{+S}$ only the diagrams 
with more than 2 external lines contribute to $\Arg D_{+{\rm in}}$. These 
lattice diagrams differ from the corresponding continuum diagrams by the 
terms of order $1/N$. Therefore for $|h_\mu|\!<\!0.5$ we have 
\begin{equation}\label{16} 
\Arg D_{+\rm in} \to \Arg D_{+C}, \qquad N\!\to\!\infty.
\end{equation}

To estimate $\Arg D_{+b0}[h_0, h_1]$ we expand the $\arctg x$ in the Taylor 
series. Keeping only the first term we replace the sum by the integral and 
substitute this expresion into the formula for $\Arg D_{+b0}$. In this way we
get
\begin{equation}\label{17}
\Arg D_{+b} = 2h_1 \arctg \frac{2h_0}{1-2h_0^2} - 2h_0 \arctg \frac{2h_1}
{1-2h_1^2}, \qquad N\!\to\!\infty.
\end{equation}
The calculations show that for all $h$ in the interval under consideration
$$
|\Arg D_{+b}| \ll |\Arg D_{+{\rm in}}|, \qquad N\!\to\!\infty.
$$

Finally, calculating $\Arg D_{+a}$ with the help of eq. (\ref{16}) we find
$$
\Arg D_{+a} = \Arg D_{+{\rm in}}, \qquad N\!\to\!\infty.
$$

\begin{figure}[t]
\epsfxsize=16cm

\vspace*{-7mm}
\hspace*{-3mm}
\rotate[r]{\epsfbox{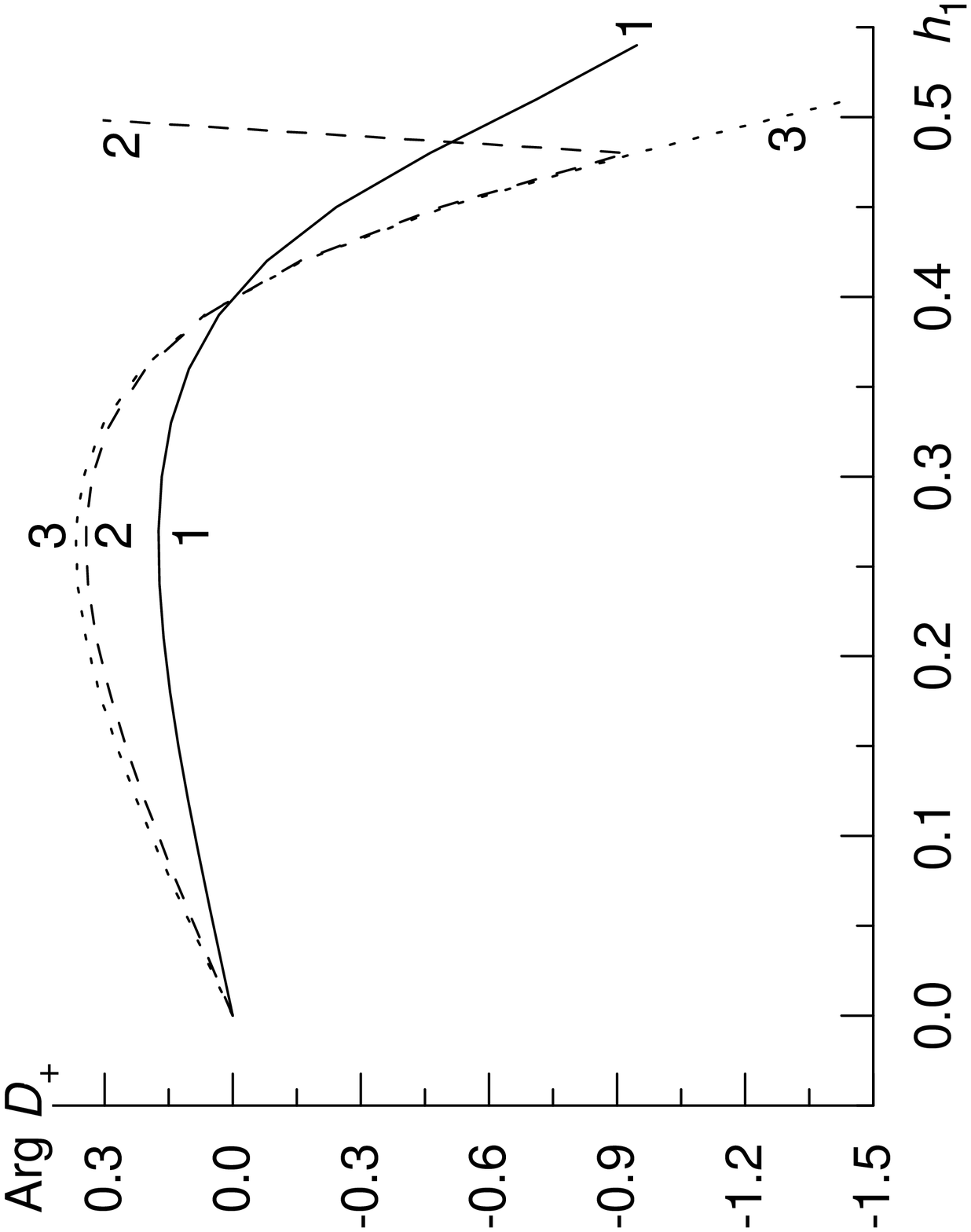}}

\vspace*{-48mm}
{\bf Fig.5.}
Arguments of positive chiral determinants $\Arg D_+$ \\
\hspace*{16mm} as functions of $h_1$ at $h_0\!=\!0.4$: \\
\hspace*{16mm} 1 -- $\Arg D_{+C}$ on the torus; \qquad  
2, 3 -- $\Arg D_{+S}$ on the lattice: \\
\hspace*{16mm} 2 -- computed by eq. (\ref{12}) at $N\!=\!32$ and 160, \\
\hspace*{16mm} 3 -- estimated by eqs (\ref{14}) -- (\ref{17}) at 
$N\!\to\!\infty$
\bigskip
\end{figure}

Our estimates show that the contribution of the edges of the Brillouin zone 
to the $\Arg D_{+S}$ is small, but the contribution of the vertices is 
approximately equal to the contribution of the interier and their sum is 
two times bigger than the continuum value:
$$
\Arg D_{+S} = 2\Arg D_{+C}.
$$

So we have here some kind of doubling phenomenon which is somewhat 
reminiscent to the phenomenon observed by Bodwin and Kovacs \cite{BodKov} in 
Rabin's formulation of Schwinger model \cite{Rab}.

Analytic estimates given above are in a good agreement with the results of 
numerical calculations presented at Fig.5. For all values of $h$ $\Arg D_{
+S}$ is two times bigger than the corresponding continuum value.

\section {Discussion}

In this paper we analyzed the U(1) model on a finite 2d lattice described by
the gauge invariant nonlocal SLAC action. It was shown that in the case of 
a constant gauge field (toron model) SLAC action generates infinite series of
divergent (in the limit $N\!\to\!\infty$) diagrams with more than two 
external lines and the value of the lattice fermion determinant does not 
agree with the known exact result for the continuum toron model. Modification 
of the SLAC action by introducing additional PV regularization cures this 
decease in the case of vectorial interaction. A minimal number of PV fields 
necessary for such regularization is equal to three.

The peculiar feature of the SLAC action on a finite lattice is large 
contribution of momenta corresponding to the vertices of the Brillouin zone.
In the case of the vectorial interaction contribution of these momenta 
results in sharp decreasing of the lattice determinant in the narrow region 
near $|h_\mu|\!=\!1/2$. However in the limit $N\!\to\!\infty$ the width of 
this region tends to zero and the model agrees with the continuum result for 
all $|h_\mu|\!<\!1/2$.

In the case of axial interaction the existence of divergent diagrams with 
more than two external lines does not allow to cancel the contribution of 
momenta $|p|\! \sim \! \frac{\pi}{a}$ by introducing PV regularization. As 
in the vectorial case the boundary momenta give a large anomalous 
contribution. As a result the phase of the lattice chiral determinant is 
two times bigger than the corresponding continuum phase.

Our results show that SLAC action supplemented by additional PV 
regularization solves successfully the problem of fermion spectrum 
doubling in the case of vectorial interaction, but fails to describe 
correctly chiral models interacting with a constant gauge field.
$$ ~ $$
{\bf Aknowlegements} \\

The authors are grateful V.Bornyakov and S.Zenkin for fruitful discussions.
This work was supported by RBRF under Grant 96-01-005511, and by INTAS Grant 
INTAS-96-370.
$$ ~ $$

\end{document}